\documentclass[
   ,final            
,sort&compress ]
  {aipproc}

\layoutstyle{6x9}

\usepackage{graphicx}
\usepackage{epsfig}
\usepackage{bm}
\newcommand{\rn}{\mathrm{n}}
\newcommand{\rP}{\mathrm{P}}
\def\ve{\varepsilon}

\newcommand{\T}{\mathsf{T}}

\newcommand{\tLN}{\mathsf{t}_{\Lambda N}}

\def\bit{\begin{itemize}}
\def\eit{\end{itemize}}
\def\bnu{\begin{enumerate}}
\def\enu{\end{enumerate}}

\def\e {{\epsilon}}

\def\nn{\nonumber }

\def\mbt{\mbox{\boldmath$\tau$}}

\def\bin#1#2{\left(\negthinspace\begin{array}{c}#1\\#2\end{array}\right)}

\def\mbpi{\mbox{\boldmath$\pi$}}

\def\M {{{\cal M}}}
\def\x{\times}
\def\Ket#1{||#1 \rangle}
\def\Bra#1{\langle #1||}
\def\lsim{\:\raisebox{-0.5ex}{$\stackrel{\textstyle<}{\sim}$}\:}

\def\ie{{\em i.e., }}

\def\nn{\nonumber }
\def\be{\begin{equation}}
\def\ee{\end{equation}}
\def\br{\begin{eqnarray}}
\def\er{\end{eqnarray}}
\def\brn{\begin{eqnarray*}}
\def\ern{\end{eqnarray*}}
\def\etc{ {\it etc}}

\def\pb {{\bf p}}
\def\Pb{ {\bf P}}
\def\rb {{\bf r}}
\def\Rb{ {\bf R}}
\def\e {{\epsilon}}
\def\lra{\leftrightarrow}
\def\mbl{\mbox{\boldmath$\lambda$}}

\def\mbt{\mbox{\boldmath$\tau$}}

\def\mbpi{\mbox{\boldmath$\pi$}}
\def\mbphi{\mbox{\boldmath$\phi$}}

\def\bra#1{\langle #1|}
\def\ket#1{|#1 \rangle}
\def\rf#1{{(\ref{#1})}}

\def\sixj#1#2#3#4#5#6{\left\{\negthinspace\begin{array}{ccc}
#1&#2&#3\\#4&#5&#6\end{array}\right\}}

\def\go{\rightarrow  }
\def\etal {\emph{et al.}}

\def\sqi{\frac{1}{\sqrt{2}}}
\def\fot{\frac{1}{2}}

\def\etc{ {\it etc}}
\def\etal{{\it et al. }}
\def\be{\begin{equation}}
\def\ee{\end{equation}}
\def\br{\begin{eqnarray}}
\def\er{\end{eqnarray}}
\def\rf#1{{(\ref{#1})}}

\def\bkU{{\rm \kern.30em\vrule width.02em height1.47ex depth-.05ex\kern-.32em U}}
\def\bkM{{\rm I\kern-.22em M}}
\def\bkPi{{\rm I\kern-.22em \Pi}}
\def\bkGa{{\rm I\kern-.17em \Gamma}}
\def\ket#1{|#1 \rangle}
\def\bra#1{\langle #1|}
\def\Ket#1{||#1 \rangle}
\def\Bra#1{\langle #1||}

\begin{document}

\title{Nonmesonic Weak Decay of $\Lambda$-Hypernuclei within Independent-Particle Shell-Model}

\author{Franjo Krmpoti\'c}{
address={Instituto de F\'isica La Plata, CONICET, \\
 Facultad de Ciencias Astron\'omicas y Geof\'isicas,\\
 Universidad Nacional de La Plata, 1900 La Plata, Argentina.},
} \iftrue

\author{Alfredo P. Gale\~ao}{
  address={Instituto de
F\'{\i}sica Te\'orica,
UNESP - Univ. Estadual Paulista,\\
Rua Dr. Bento Teobaldo Ferraz 271 - Bl. II, 01140-070 S\~ao Paulo,
SP, Brazil.},
}

\author{Mahir S. Hussein}{
  address={Departamento de F\'isica Matem\'atica,\\ Instituto
de F\'isica da Universidade de S\~ao Paulo,\\
Caixa Postal 66318, 05315-970 S\~ao Paulo, SP, Brazil},
}

\fi
\begin{abstract}
After a short introduction to  the nonmesonic weak decay (NMWD)
$\Lambda N\go nN$ of $\Lambda$-hypernuclei we discuss the
long-standing puzzle on the ratio $\Gamma_n/\Gamma_p$, and some
recent  experimental evidences that signalized towards  its final
solution. Two versions of the  Independent-Particle-Shell-Model
(IPSM) are employed to account for the nuclear structure of  the
final residual nuclei. They are: (a) IPSM-a, where  no
correlation, except for the Pauli principle, is  taken into
account, and (b) IPSM-b, where the highly excited  hole states are
considered to be quasi-stationary and are described by
Breit-Wigner distributions, whose widths are estimated from the
experimental data. We evaluate  the  coincidence spectra in
$^{4}_\Lambda$He, $^{5}_\Lambda$He, $^{12}_\Lambda$C,
$^{16}_\Lambda$O, and $^{28}_\Lambda$Si, as a function of the  sum
of kinetic energies $E_{nN}=E_n+E_N$ for $N=n,p$.  The recent
Brookhaven National Laboratory experiment E788 on $^4_\Lambda$He,
is interpreted within the IPSM . We found that  the shapes of all
the spectra are basically tailored by the kinematics of the
corresponding phase space, depending very weakly on the dynamics,
which is gauged here by the one-meson-exchange-potential. In spite
of the straightforwardness of the approach a good agreement with
data is achieved. This might be an indication that the
final-state-interactions and the two-nucleon induced processes are
not very important  in the decay of this hypernucleus. We have
also found that the $\pi+K$ exchange potential with soft
vertex-form-factor cutoffs $(\Lambda_\pi\approx 0.7$ GeV,
$\Lambda_K\approx 0.9$ GeV), is able to account simultaneously for
the available experimental data related to $\Gamma_p$ and
$\Gamma_n$ for  $^4_\Lambda$H, $^4_\Lambda$He, and $^5_\Lambda$He.
\end{abstract}

\maketitle

\section{Introduction}
 The nonmesonic weak decay   (NMWD)  of $\Lambda$ hypernuclei, $\Lambda
N\go nN$ ($N=p,n$), takes place only within nuclear  environment.
Without producing any additional on-shell particle (as does  the
mesonic weak decay  $\Lambda \go \pi N$)
 the mass is changed by $176$ MeV,
and the strangeness by $\Delta {\sf S}=1$, which implies that we
are witnessing the most drastic metamorphosis of an elementary
particle within the nucleus. As such, the hypernuclei can be
considered as a powerful "laboratory" for unique investigations of
baryon-baryon strangeness- changing weak interactions, and the
NMWD could play an important role in the stability of rotating
neutron stars with respect to gravitational wave
emission~\cite{Da04, Ju08}.

Same as the  free $\Lambda$ hyperon, they are mostly produced via
the strong interactions, \ie in the reaction processes
$\pi^+n\go\Lambda K^+$, $K^-n\go\pi^-\Lambda$ and
$K^-p\go\pi^0\Lambda$, by making use of the pion ($\pi$) and kaon
($K$) beams. They also
 basically decay  through
the weak interactions, as the free $\Lambda$ does. Yet, as it is
well known and explained below, there are some very important
differences in the corresponding decaying modes.
\begin{figure}[ht]
\begin{tabular}{cc}
\includegraphics[width=0.4\linewidth,clip=]{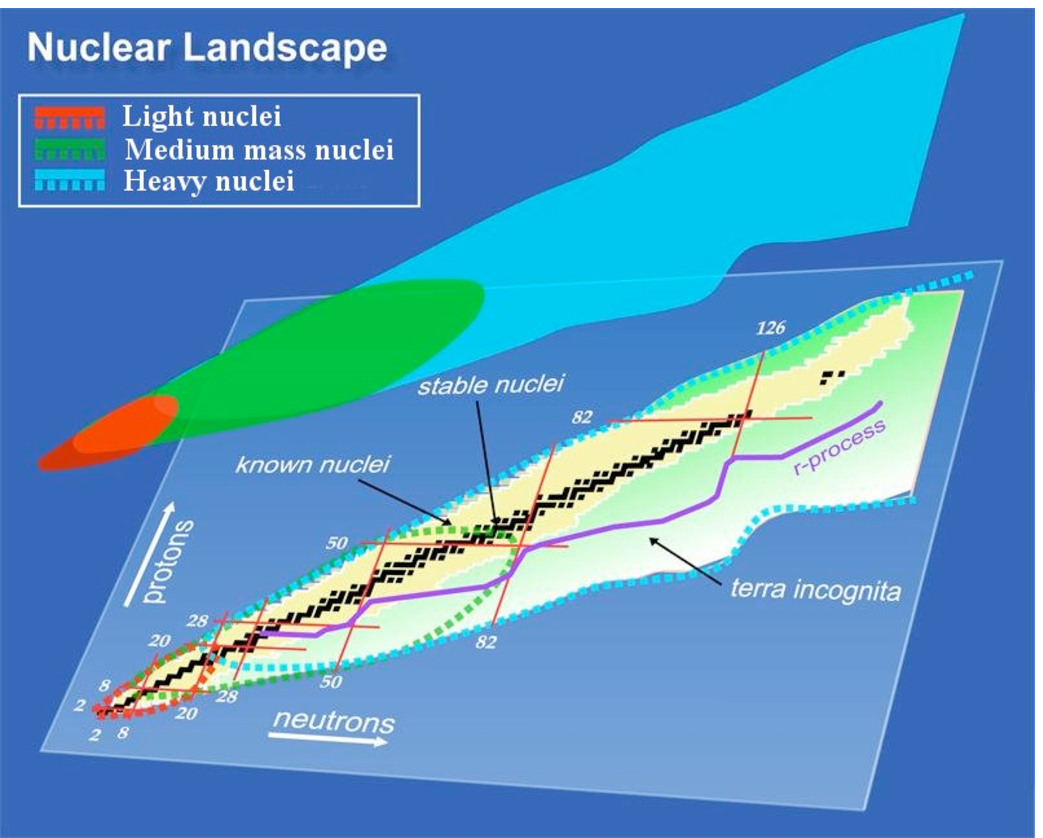} &
\includegraphics[width=0.4\linewidth,clip=]{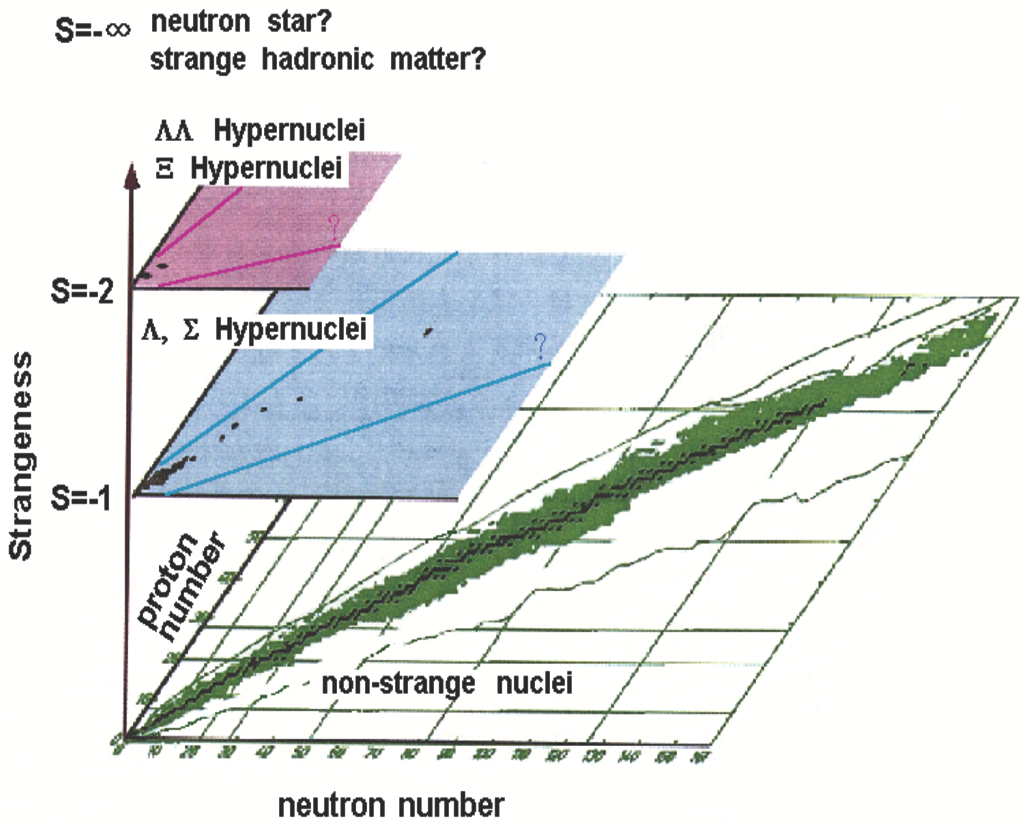}
\end{tabular}
\caption{\label{F1} On the left and right panels are represented,
respectively, the standard nuclear chart of nuclides, and the
hypernuclear chart of isotopes. The NMWD corresponds to a jump
from the surface $S=1$ in the three-dimensional chart $({\sf
N,Z,S})$ to the standard $({\sf N,Z})$ chart.}
\end{figure}

\begin{figure}[ht]
\includegraphics[width=0.6\linewidth,clip=]{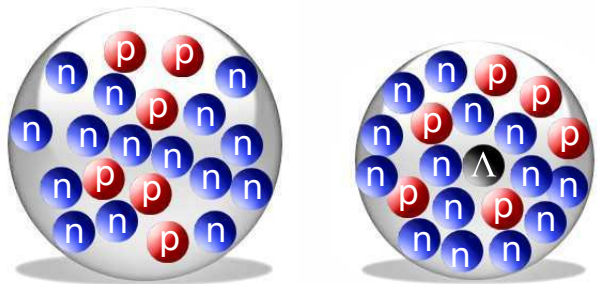}
\caption{\label{F2} Pictorial representation of shrinkage of
$^{20}$C nucleus by the addition of an $\Lambda$-hyperon to build
up the $^{21}_\Lambda$C
 hypernucleus.}
\end{figure}

Not less important  is the fact that with the incorporation of
strangeness, the radioactivity domain is  extended to three
dimensions $({\sf N,Z,S})$, as illustrated in Fig. \ref{F1}.
 The best studied systems are nuclei containing a $\Lambda$-hyperon, which
because of the additional binding  are even richer in elements
than the ordinary $({\sf N,Z})$ domain.
  (For instance, while the one-neutron separation energy in $^{20}$C is $1.01$ MeV,
   it is $1.63$ MeV in $^{21}_\Lambda$C ~\cite{Sa08}.)
 This attribute of hypernuclei
has motivated a recent proposal to  produce  neutron rich
$\Lambda$-hypernuclei at J-PARC, including
$^{9}_\Lambda$He~\cite{Sa09}. The shrinkage of  the $^{20}$C
nucleus by the addition of an $\Lambda$-hyperon to build up the
$^{21}_\Lambda$C
 hypernucleus is illustrated in Fig. \ref{F2}.

\begin{figure}[th]
\begin{tabular}{ccc}
\includegraphics[width=0.5\linewidth,clip=]{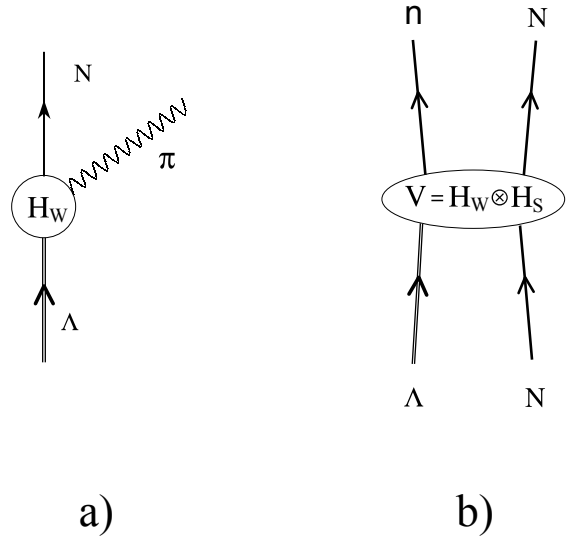} &~~~~~~~~~~~~&
\end{tabular}
\caption{\label{F3} Schematic representation of: a) mesonic
(nonleptonic) decay $ \Lambda \go N+\pi$
 induced by the weak vertex $H_W$, and b)  nonmesonic weak decay $ \Lambda N\go nN$,
  engendered by the product of a  weak vertex $H_W$ and a strong vertex $H_S$.}
\end{figure}

The  free $\Lambda$ hyperon decays (as represented schematically
by the first graph in Fig. \ref{F3}) nearly $100$ \% of the time
by the $\Lambda\go N\pi$ weak-mesonic mode \brn
 \Lambda\go \left\{\begin{array}{ccc}
 p+\pi^{-} && (64.1\%) \\
         n+\pi^{0} && (35.7\%),
\end{array}\right.
\ern with the total transition rate
$\Gamma^0_{\pi^-}+\Gamma^0_{\pi^0}=\Gamma^0= 2.50 \cdot 10^{-6}$
eV (which corresponds to the lifetime $\tau^0= 2.63 \cdot
10^{-10}$ sec). For the decay at rest  the energy-momentum
conservation implies \brn
M_\Lambda=M_N+\frac{p_N^{2}}{2M_N}+\sqrt{p_\pi^{2}+m_\pi
^{2}};~~~~p_N&\equiv&p_\pi. \ern Therefore the energy released is
\brn Q_0=M_\Lambda-M_N-m_\pi\cong 37~ \mbox{MeV}, \ern and the
kinetic energies and momenta in the final state are: \brn
T_N&=&\frac{(M_\Lambda-M_N) ^{2}+m_\pi ^{2}}{2M_\Lambda}\cong 5~
\mbox{MeV};
\nn\\
T_\pi&=&Q_0-T_N\cong 32~ \mbox{MeV},\nn\\
p_N&\equiv&p_\pi=\sqrt{(T_N+M_N)^2-M_N^2}\cong 100~ \mbox{MeV}/c.
\ern During the decay
  the isospin is changed by $\Delta T=1/2$ and $3/2$
and its projection by $\Delta M_T=-1/2$. However, as  the above
experimental data can be accounted for fairly well by neglecting
the  $\Delta T=3/2$ component, one end up with $\Delta T=1/2$
rule, which leads to the estimate
$\Gamma_{\pi^-}/\Gamma_{\pi^0}=2$, while the experimental result
is $64.1/35.7=1.80$.

Assuming the $\Delta T=1/2$ rule, the phenomelogical weak
Hamiltonian for the process depicted in Fig. \ref{F3}1 can be
expressed as: \br
H_W&=&-iG_Fm_\pi^2\overline{\psi}_N\left(A_\pi+B_\pi\gamma_5\right)
\mbphi_\pi\cdot\mbt\psi_\Lambda\bin{0}{1}, \label{1.1}\er where
$G_Fm_\pi^2=2.21\x 10^{-7}$ is the weak coupling constant. The
empirical constants $A_\pi=1.05$  and $B_\pi=-7.15$, adjusted to
the observables of the free $\Lambda$ decay, determine the
strengths of parity violating and parity conserving amplitudes,
respectively. The nucleon, $\Lambda$ and pion fields are given by
$\psi_N$ and  $\psi_\Lambda$ and $\mbphi_\pi$, respectively, while
the isospin spurion $\bin{0}{1}$ is included in order to enforce
the empirical $\Delta T={ 1/2}$ rule.

 The free $\Lambda$ hyperon weak decay  is radically modified  in the nuclear
environment because the nucleon and  the hyperon now move,
respectively,
  in  the mean fields $U_N$ and $U_\Lambda$, which come from the $NN$ and $N\Lambda$
  interactions. $U_N$ and $U_\Lambda$ are characterized by the single particle energies
(s.p.e.)  $\varepsilon_N$ and $\varepsilon_\Lambda$, and we have
to differentiate between:

$\bullet$ {\it Mesonic Weak Decay (MWD)}:
 The basic process is again represented by the first graph shown in Fig. \ref{F3},
 and  described by the hamiltonian \rf{1.1}. Yet,
the energy-momentum conservation is different: \br
M_\Lambda=M_N-\varepsilon_\Lambda +\varepsilon_N^\uparrow
+\frac{p_\pi^{2}}{2(A-1)M_N}+\sqrt{p_\pi^{2}+m_\pi ^{2}},
\label{1.2}\er where $A$ is the mass number, and
$\varepsilon_N^\uparrow$ are the s.p.e. of the loosely bound
states above the Fermi energy $\varepsilon_{N}^{F}$. They are of
the order of a few MeV, while $\varepsilon_\Lambda$ is the energy
of  the $0s_{1/2}$ state and goes from $-11.7$~MeV for
$^{13}_\Lambda $C to $-26.5$~MeV for $^{208}_\Lambda$ Pb
\cite{Us99}. Thus, the corresponding Q-values \br
Q_M=M_\Lambda-M_N-m_\pi+\varepsilon_\Lambda
-\varepsilon_N^\uparrow, \label{1.3}\er are significantly smaller
than $Q_0$, particularly for medium and heavy nuclei. This small
value of $Q_M$ makes, as illustrated in \cite[Fig. 2]{Kr03},  the
MWD  to be  hindered due to the Pauli principle. In fact, the
experimental decay rates
$\Gamma_{\pi^-}+\Gamma_{\pi^0}=\Gamma_M\equiv \Gamma_M
(\Lambda\rightarrow N\pi$)
 are of the order of
$\Gamma^0$ only for  nuclei with $A\le 4$, and they rapidly fall
as a function of nuclear mass. For instance, in $^{12}_\Lambda$ C:
$\Gamma_{\pi^0}/\Gamma^0=0.217\pm 0.084$ and
$\Gamma_{\pi^-}/\Gamma^0=0.052^{+0.063}_{-0.035}$. (For a recent
theoretical study of the MWD see Ref.~\cite{Ga09}.)
\begin{figure}
\includegraphics[width=0.6\linewidth,clip=]{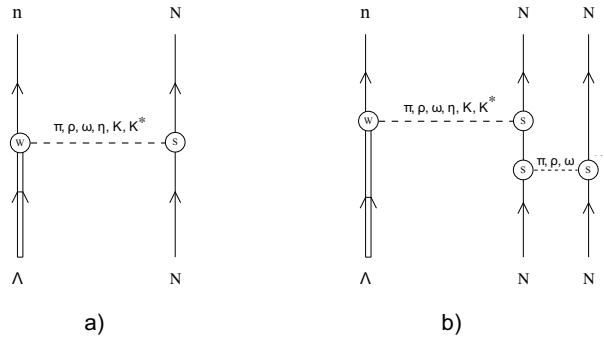}
\caption{\label{F4} a)  One-nucleon, and  b) two-nucleon induced
$\Lambda$-decay in nuclei. Strictly speaking the OME diagrams are
only valid for nonstrange-mesons $\pi,\rho,\omega$, and $\eta$.
For strange-mesons $K$, and $K^*$ the vertices $W$, and $S$ should
be exchanged, \ie $W\go S$, and $S\go W$. }
\end{figure}

$\bullet$ {\it Nonmesonic Weak Decay (NMWD)}: New nonmesonic
decay  channels $\Lambda N\rightarrow nN$
 become open inside the nucleus, where there are no
pions in the final state; it is represented schematically by the
second diagram in Fig. \ref{F3}. The corresponding transition
rates can be stimulated either by protons,
$\Gamma_p\equiv\Gamma(\Lambda p\rightarrow np)$, or by neutrons,
$\Gamma_n\equiv\Gamma(\Lambda n\rightarrow nn)$. The
energy-momentum conservation and the Q-value are, respectively:
\br M_\Lambda&=&M_N-\varepsilon_\Lambda -\varepsilon_N^\downarrow
+\frac{p_n^{2}}{2M_N}+\frac{p_N^{2}}{2M_N}+\frac{|\pb_n+\pb_N|^2}{2M_N(A-2)},
\label{1.4}\er and \br Q_{NM}=M_\Lambda-M_N+\varepsilon_\Lambda
+\varepsilon_N^\downarrow. \label{1.5}\er

Since  the mean energy  of the bound single-particle states is
$\varepsilon_N^\downarrow\sim -30$ MeV, the Q-value is $Q_{NM}\sim
120-135$ MeV, and this is basically the kinetic energy of the two
particles that are ejected from the hypernucleus. Therefore, the
NMWD  possesses a large phase space in the continuum, as
illustrated in \cite[Fig. 3]{Kr03}, and the momenta $\pb_n$,  and
$\pb_N$ of  two outgoing nucleons are relatively large ($\sim 420$
MeV). Therefore, the non-mesonic mode is not blocked by the Pauli
principle, and  dominates over the mesonic mode for all but the
$s$-shell hypernuclei.

It is assumed very often that the  hypernuclear NMWD $\Lambda
N\rightarrow nN$
 is triggered via the exchange of a virtual meson, and the obvious candidate
 is the one-pion-exchange (OPE) mechanism, where the strong  Hamiltonian
\br
H_S^{\pi}&=&ig_{NN\pi}\bar{\psi}_N\gamma_5\mbpi\cdot\mbt\psi_N,
\label{1.6}\er with $g_{NN\pi}=13.4$, accompanies the weak
Hamiltonian \rf{1.1}~\cite{Ad67}.
\begin{figure}
\includegraphics[width=0.8\linewidth,clip=]{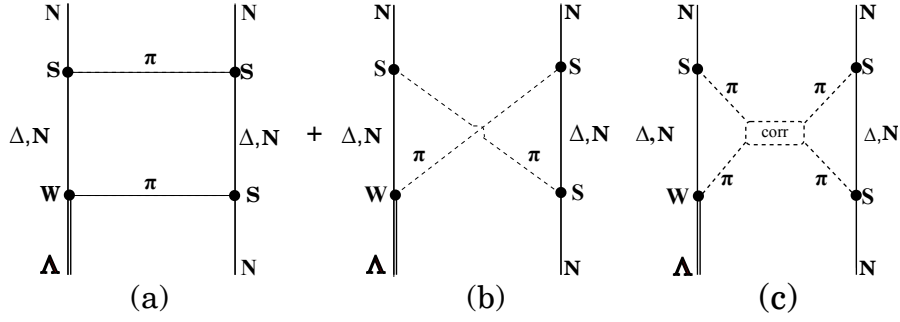}
\caption{\label{F5} (a)  direct, and (b) crossed uncorrelated, and
(c) correlated two-pion-exchange diagrams for the weak $\Lambda
N\go nN$ transition amplitude.}
\end{figure}

Later on, the full  one meson-exchange (OME) model has been
introduced by Dubach \etal~\cite{Du96}, as schematically
represented by the first graph in Fig. \ref{F4}. Also are
considered frequently the two-nucleon induced NMWD, represented by
the second diagram in in Fig. \ref{F4}. Here, one or two bound
nucleons are expelled to the continuum by the nuclear ground state
correlations, and one of them, together with the hyperon
$\Lambda$, exchanges one meson giving rise to three decaying
nucleons, \ie $\Lambda nN\rightarrow nNN$. The corresponding decay
rate is denoted as $\Gamma_2$, and the total weak decay rate of a
$\Lambda$-hypernucleus is then: \br \Gamma_T = \Gamma_M +
\Gamma_{NM}, \label{1.7}\er where: \br \Gamma_M = \Gamma_{\pi^-}+
\Gamma_{\pi^0},~~ \Gamma_{NM}= \Gamma_1 + \Gamma_2,~~ \Gamma_1 =
\Gamma_n + \Gamma_p. \label{1.8}\er The OME potential is sometimes
complemented with the contributions of uncorrelated ($2\pi$) and
correlated ($2\pi/\sigma$) two-pion-exchange~\cite{Ch07}, which
are illustrated in  see Fig. \ref{F5}.

\section {$\Gamma_n/\Gamma_p$ puzzle}

\begin{figure}[h]
\includegraphics[width=0.65\linewidth,clip=]{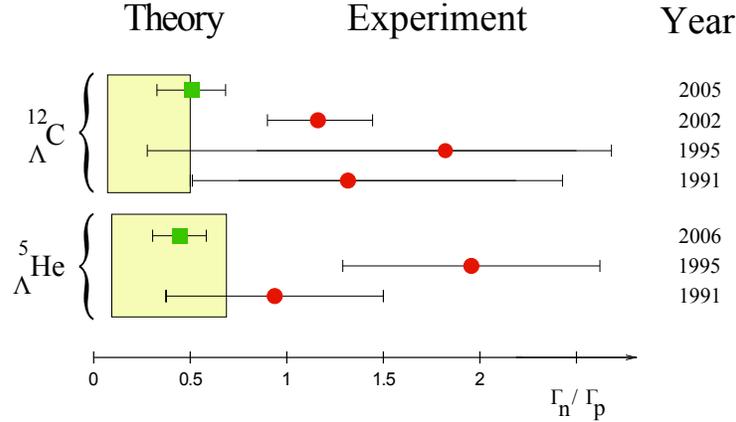}
\caption{\label{F6} $\Gamma_n/\Gamma_p$ puzzle: The  experimental
values of the ratio $\Gamma_{n/p}$ are: a) $^{5}_\Lambda $He
$\Gamma_{n/p}=0.93\pm 0.55$ (BNL)~\cite{Sz91}, $1.97\pm 0.67$
(KEK)~\cite{No95}, $0.45\pm0.11\pm0.03$ (KEK-E462)~\cite{Ka06},
and b) $^{12}_\Lambda $C: $\Gamma_{n/p}=1.33^{+1.12}_{-0.81}$
(BNL)~\cite{Sz91}, $1.87\pm 0.59^{+0.32}_{-1.00}$
(KEK)~\cite{No95}, $1.17^{+0.09+0.20}_{-0.08-0.18}$
(KEK)~\cite{Ha02}, $0.51\pm 0.13\pm0.05$ (KEK)~\cite{Ki03,Ki06},
while  the theoretical estimates for $\Gamma_{n/p}$ vary between
$0.09$ and $0.70$ for $^{5}_\Lambda $He, and between $0.08$ and
$0.50$ for $^{12}_\Lambda $C. }
\end{figure}

Large experimental values
of the ratio $\Gamma_{n/p}=\Gamma_n/\Gamma_p$ in $^{5}_\Lambda $He
 and $^{12}_\Lambda $C, measured before the year $2003$~\cite{Sz91,No95,Ha02},
  were a cumbersome puzzle for
 the theorists during almost  two  decades, as schematically represented in Fig. \ref{F6}.
 In fact,
following the pioneering investigations of Adams \cite{Ad67}
several calculations  have been done within OPE coupling scheme of
the total NMWD rate, and  the ratio
$\Gamma_{n/p}=\Gamma_n/\Gamma_p$
reproducing reasonably well the first one, but failing badly
for the second  observable. (see Refs.~\cite{Os98,Al02,Al04,Pa07,Ch08}, and references therein).

The deficiency of the OPE model was attributed to effects of short
range physics, which should be quite important in view of the
large momentum transfers involved. Although there have been some
attempts to account for this fact by making use of quark models to
compute the shortest range part of the transition
potential~\cite{Che83,Ma94,In96,In98,Sa00}, most of the
theoretical work opted for the addition of other, heavier mesons
in the exchange
process~\cite{Mc84,Du96,Ta85,Na88,Pa95a,Ra91,Ra92,Pa95,Pa97,Ha01,Pa01,Ba02,Sa02,Ba03,
Bau03,Ba05,Kr03}. None of these models gives fully satisfactory
results.
 Inclusion of correlated two-pion exchange has not been completely successful either~\cite{Sh94,It02}.
 Nor have the
addition of uncorrelated two-pion exchange, two-nucleon induced transitions or medium effects,
treated within the nonrelativistic~\cite{Os85,Al91,Ra94,Ra97a,Al02,Ji01}
or relativistic~\cite{Zh99} propagator approaches, been of
much help.

 Yet, several important experimental advances  in NMWD
 have been made in recent years,
which have allowed to establish more precise values of the
neutron- and proton-induced  transition rates  $
\Gamma_n$
 and $\Gamma_p$,
solving in this way the long-standing puzzle of
the branching ratio  $\Gamma_{n/p}$.
They are: 1) the new
high quality measurements of  single-nucleon spectra $S_{N}(E)$,
as a function of one-nucleon energy $E_N\equiv E$ done in
Refs.~\cite{Ki03,Ok04,Par07,Ag08},
and 2) the first  measurements of the two-particle-coincidence spectra as a
function of the sum of  kinetic energies $E_n+E_N\equiv E$, $S_{nN}(E)$, and of
the opening angle $\theta_{nN}\equiv\theta$, $S_{nN}(\cos\theta)$, done in
Refs.~\cite{Ok05,Ou05,Ka06,Ki06,Bh07,Par07}.

\section{Transition Rates}\label{Sec3}
 To derive the NMWD rate we start from the Fermi Golden Rule.
For  a hypernucleus (in its ground state with spin
$J_I$ and energy $E_{J_I}$) to residual nuclei (in the several
allowed states with spins $J_F$ and energies $E_{\alpha_NJ_F}$)
and two free nucleons $nN$ (with  total spin $S$ and total kinetic
energy $E_{nN}=E_n+E_{N}$), reads
\begin{eqnarray}
\Gamma_N &=& 2\pi \sum_{S\alpha_NJ_F}
\int |\bra{\pb_n\pb_N S;\alpha_NJ_F}V\ket{J_I}|^2
 \delta(\Delta_{\alpha_NJ_F}-E_R-E_{nN}) \frac{d{\bf
p}_n}{(2\pi)^3}\frac{d{\bf p}_N}{(2\pi)^3}, \label{3.1}
\end{eqnarray}
where for the sake of simplicity we have suppressed the magnetic quantum numbers.
The NMWD dynamics, contained within the  weak hypernuclear
transition potential $V$, will be described  by the OME
model, whose most commonly used version  includes the exchange
of the full pseudoscalar ($\pi, K, \eta$) and vector
($\rho,\omega,K^*$) meson octets (PSVE), with the weak coupling
constants obtained from soft meson theorems and
$SU(6)_W$~\cite{Pa97,Du96}.
The wave functions for the kets
$\ket{\pb_n\pb_N SM_SJ_FM_F}$ and $\ket{J_IM_I}$ are assumed to be
antisymmetrized and normalized, and the two emitted nucleons $n$
and $N$ are described by plane waves.  Initial and final short
range correlations are included phenomenologically at a simple
Jastrow-like level, while the finite nucleon size effects at the
interaction vertices are gauged by monopole form
factors~\cite{Pa97,Ba02}. Moreover,
 \be
 E_R =\frac{|\bm{p}_n+
\bm{p}_N|^2}{2M(A-2)}= \frac{E_{nN} + 2\cos\theta_{nN} \sqrt{E_n
E_N}}{A - 2},
 \label{3.2}\ee
 is the
recoil energy of the residual nucleus, and
\begin{equation}
\Delta_{\alpha_NJ_F}=\Delta M+E_{J_I}-E_{\alpha_NJ_F},
\hspace{1cm} \mbox{with} \hspace{1cm}
\Delta M=M_\Lambda-M=176~~ \mbox {MeV},
\label{3.3}
\end{equation}
is the liberated energy.

It could be convenient to perform a transformation to the  relative  and
c.m. momenta ($\pb=\fot(\pb_n-\pb_N $),  $\Pb=\pb_n+\pb_N$), coordinates
($\rb=\rb_n-\rb_N$,  $\Rb=\fot(\rb_n+\rb_N)$) and orbital angular momenta
$\bm{l}$ and $\bm{L}$,
and to express the energy conservation as
\be
E_{nN}+E_r-\Delta_{\alpha_NJ_F}=\e_p+\e_P-\Delta_{\alpha_NJ_F}=0,
\label{3.4}\ee
where
\be
\e_p=\frac{p^2}{M},
\hspace{1cm}E_r=\frac{P^2}{2M(A-2)},
\hspace{1cm}\e_P=\frac{P^2}{4M}\frac{A}{A-2}=\frac{A}{2}E_r,
\label{3.5}\ee
are, respectively, the energies of  the relative motion of the outgoing pair,
of the recoil, and of the total c.m. motion (including the recoil).

Following  the analytical developments
done in Ref.~\cite{Ba02},   the transition rate can be expressed
as a function of the c.m. energy $\e_P$:
 \br
\Gamma_{N}&=&
\frac{16M^3}{\pi}\left(\frac{A-2}{A}\right)^{3/2}\int_{0}^{\Delta} d\e_P
\sum_{\alpha_NJ_F}
\sqrt{\e_P(\Delta_{\alpha_NJ_F}-\e_P)}\mathcal{F}_{\alpha_NJ_F}(pP).
\label{3.6}\er
 It is  understood that the square root
should be replaced by zero whenever its
argument is negative. Here
\br
\mathcal{F}_{\alpha_NJ_F}(pP) &=&\hat{J}_I^{-2}\sum_{S\lambda lLTJ}
\left|\sum_{j_N} \M(plPL\lambda SJ\T;j_\Lambda j_N J\tLN)
\Bra{J_I}\left( a_{j_N}^\dag a_{j_\Lambda}^\dag\right)_{J}
\Ket{\alpha_NJ_F}\right|^2,
\nn\\
 \label{3.7}\er
and
\begin{eqnarray}
\M(plPL\lambda SJ\T;j_\Lambda j_N J\tLN)
&=&\sqi\left[1-(-)^{l+S+T}\right]
\nonumber \\ &\times&
({plPL\lambda SJ\T}|V|{j_\Lambda j_N J\tLN}).
\label{3.8}\er
$\T\equiv \{TM_T,M_T=m_{t_\Lambda}+m_{t_N}\}$, $\tLN\equiv \{t_\Lambda=1/2,m_{t_\Lambda}=-1/2,
t_N=1/2,m_{t_N}\}$, with
$m_{t_p}=1/2$,  and $m_{t_n}=-1/2$, and $l$ and $L$ stand for quantum numbers of the relative
and c.m. orbital angular momenta in the $\Lambda N$ system.
The transition matrix elements $\M(plPL\lambda SJ\T;j_\Lambda j_N J\tLN)$ depend on the  c.m. and relative  momenta, which
are given in terms of the integration variable $\e_P$ by
 \be
P=2\sqrt{\frac{A-2}{A}M\e_P},\hspace{0.5cm}p=\sqrt{M (\Delta_{\alpha_NJ_F}-\e_P)},
 \label{3.9}\ee
where the energy conservation condition has been used.
The angular momentum couplings ${\bf l}+{\bf L}={\mbl}$,
and ${\mbl}+{\bf S}={\bf J}$ have been carried out, $\hat{J}\equiv\sqrt{2J+1}$,
and $A=Z+N+1$ is the total number of baryons.

It is self-evident that for $A\go \infty$ one obtains the same
result as in Refs. \cite{Ba02,Kr03,Ba03}. It is also worth noting
that the overall outcome of the recoil on $\Gamma_N$ is very small,
mostly because the effect of  the factor
$\left(\frac{A-2}{A}\right)^{3/2}$ in Eq. \rf{3.6} is, to a great
extent, cancelled by the effect of  the factor
$\left(\frac{A}{A-2}\right)^{3/2}$ originating from
$\sqrt{\e_P(\Delta_{\alpha_NJ_F}-\e_P)} d\epsilon_P$. This is the
reason why we have not included the recoil previously.

 From the relation
\be
E_{nN}=\Delta_{\alpha_NJ_F}-\frac{2}{A}\e_P,
\label{3.10}\ee
which  follows from \rf{3.4} and \rf{3.5}, one can now
easily derive the  spectrum  of $\Gamma_N$ as a function of the sum energy $E_{nN}$~\cite{Ba08}:
\begin{equation}
\Gamma_N =
\frac{4M^3}{\pi}\sqrt{A(A-2)^3}\int_0^\Delta  dE_{nN}
\sum_{\alpha_NJ_F}
\sqrt{(\Delta_{\alpha_NJ_F}-E_{nN})(E_{nN}-\Delta_{\alpha_NJ_F}')},
\mathcal{F}_{\alpha_NJ_F}(pP).
\label{3.11}\ee
where
\br
p&=&\sqrt{\frac{MA}{2}\left(E-\Delta_{\alpha_NJ_F}'\right)},
\nn\\
{P}&=&\sqrt{2M(A-2)(\Delta_{\alpha_NJ_F}-E)},
\label{3.12}\er
\be
\Delta_{\alpha_NJ_F}'=\Delta_{\alpha_NJ_F}\frac{A-2}{A},
\label{3.13}\ee
and the condition
\be
\Delta_{\alpha_NJ_F}'\le E\le \Delta_{\alpha_NJ_F},
\label{3.14}\ee
has to be fulfilled for each contribution.

In the same way from \rf{3.6}, and \rf{3.7} we can easily arrive to an expression
for $\Gamma_{N}$ as an integral on  the c.m. momentum $P$, namely
  \br
\Gamma_{N}&=&
\frac{2M}{\pi}\sqrt{\frac{A-2}{A}}\int_{0}^{P_\Delta}dP
\sum_{\alpha_NJ_F}
P^2\sqrt{P^2_{\Delta_{\alpha_NJ_F}}-P^2}\mathcal{F}_{\alpha_NJ_F}(pP),
\label{3.15}\er
with
  \be
p=\fot\sqrt{\frac{A}{A-2}}\sqrt{P^2_{\Delta_{\alpha_NJ_F}}-P^2},\hspace{0.5cm}
P_{\Delta_{\alpha_NJ_F}}=2\sqrt{\frac{A-2}{A}\Delta_{\alpha_NJ_F}},
\label{3.16}\ee
and similarly for $P_\Delta$. t is clear that  the condition
$P\le P_{\Delta_{\alpha_NJ_F}}$ has to be fulfilled for each contribution.

Following step by step the developments done in Refs.
\cite{Ba05,Ba07,Ba08}, the Eq.
\rf{3.1} can be cast in the form
\br
\Gamma_N&=&\frac{4}{\pi}\sum_{\alpha_NJ_F}\int
d\cos\theta \int p_N^2dp_N \int p_n^2dp_n\,  \delta(\Delta_{\alpha_NJ_F}-E_R-E_{nN})
\mathcal{F}_{\alpha_NJ_F}(pP),
 \label{3.17}\er
where the  c.m. and relative  momenta, given in terms of the integration variables in \rf{3.17}
 read
 \br
{P}&=&\sqrt{(A-2)(2M\Delta_{\alpha_NJ_F}- p_n^2 -p_N^2)},
 \label{3.18}\er
and
 \br
p&=&\sqrt{M \Delta_{\alpha_NJ_F}- \frac{A}{4(A-2)} P^2}.
 \label{3.19}\er

Next, the $\delta$ function in \rf{3.17} can be put in the form
 \br
\frac{A-2}{A-1}\frac{2M}{|p_n^+-p_n^-|}\left[\delta(p_n-p_n^+)+\delta(p_n-p_n^-)\right],
\label{3.20}\er
 where
 \br
 p_n^\pm
&=&-(A-1)^{-1}\left[p_N\cos{\theta_{nN}}\mp
\sqrt{2\emph{}M(A-2)(A-1)\Delta_{\alpha_NJ_F}-p_N^2\left[(A-1)^2-\cos^2{\theta_{nN}}\right]}\right].
\nn\\
\label{3.21}\er
The Eq. \rf{3.1} becomes now
 \br
\Gamma_N&=&\frac{8M}{\pi}\frac{A-2}{A-1}\sum_{\alpha_NJ_F} \int _{-1}^{+1}d
\cos\theta_{nN}\int p_N^2 dp_N\frac{(p_n^+)^2}{|p_n^+-p_n^-|}
\mathcal{F}_{\alpha_NJ_F}(pP)
_{p_n \to p_n^+}
\nn\\
&+& ( p_n^+\lra p_n^-),
\label{3.22}\er
where the notation ${p_n \to p_n^+}$ indicates that
$\M(plPL\lambda SJ\T;j_\Lambda j_N J\tLN)$ is to be computed with
$P$ and $p$ given by Eqs.~\rf{3.18} and \rf{3.19} with $p_n$ replaced by
$p_n^+$. We have shown numerically that the last term in \rf{3.25}
is negligibly small in comparison with the first one, and therefore
it will be omitted  from now on.

 With the simple change of variable $p\go \sqrt{2ME}$ one finally gets
\br
\Gamma_N&=&(A-2)\frac{8M^3}{\pi}\sum_{\alpha_NJ_F}  \int _{-1}^{+1}d \cos\theta_{nN}
\int_0^{E_{\alpha_NJ_F}^{max}} dE_{N}
\sqrt{\frac{E_{N}}{E'}}E_n\mathcal{F}_{\alpha_NJ_F}(pP),
 \label{3.23}\er
where 
 \br
E'&=&(A-2)(A-1)\Delta_{\alpha_NJ_F}-E_{N} \left[(A-1)^2-\cos^2\theta_{nN}\right],
\label{3.24}\er
  \br
  E_n&=&\left[
\sqrt{E'} -\sqrt{E_{N}}\cos{\theta_{nN}}\right]^2(A-1)^{-2},
\label{3.25}\er
 \br
{P}&=&\sqrt{2M(A-2)(\Delta_{\alpha_NJ_F}- E_n -E_N)},
 \label{3.26}\er
and
 \br
p&=&\sqrt{M \Delta_{\alpha_NJ_F}- \frac{A}{4(A-2)} P^2},
 \label{3.27}\er
 It might be worth noticing that, while $E'$ does not have a direct physical
meaning, $E_n$ is the energy of the neutron that is
the decay partner of the nucleon $N$ with energy $E_N$.  The
maximum energy of integration in \rf{3.23} is
\be
E^{max}_{\alpha_NJ_F}=\frac{A-2}{A-1}\Delta_{\alpha_NJ_F}.
 \label{3.28}\ee

\section{Independent-Particle Shell-Model}
The formulas derived so far for the NMWD rates $\Gamma_N$ are totally general, and
don't depend on the nuclear model that is used to describe the initial
hypernuclear state $\ket{J_I}$ and the final nuclear states $\ket{\alpha_NJ_F}$.
Beneath we describe the Independent-Particle Shell-Model (IPSM), which  is widely
used in finite nuclei.
As usually done it will be assumed that the
hyperon in the state ${j_\Lambda}$, with  single-particle energy
$\e_{j_\Lambda}$,  is  weakly coupled to the $A-1$ core, with spin
${J_C}$ and energy $E_C=E_{J_I}-\e_{j_\Lambda}$. Then the initial
state is $\ket{J_I}\equiv\ket{(J_Cj_\Lambda)J_I}$, and

\begin{equation}
\Delta_{\alpha_NJ_F}=\Delta M+E_C+\e_{j_\Lambda}-E_{\alpha_NJ_F},
\hspace{1cm} \mbox{with} \hspace{1cm}E_C=E_{J_I}-\e_{j_\Lambda}
\label{4.1}
\end{equation}
Moreover,  the
spectroscopic amplitude  in Eq. \rf{3.7} can be rewritten
as
 \br
\Bra{J_I}\left( a_{j_N}^\dag
a_{j_\Lambda}^\dag\right)_{J}\Ket{\alpha_NJ_F}
&=&(-)^{J_F+J+J_I}\hat{J}\hat{J}_I
\sixj{J_C}{J_I}{j_\Lambda}{J}{j_N}{J_F}
\Bra{J_C}a_{j_N}^\dag\Ket{\alpha_NJ_F}. \label{4.2}\er
\begin{figure}[th]
{\includegraphics[width=10.cm,height=6.cm]{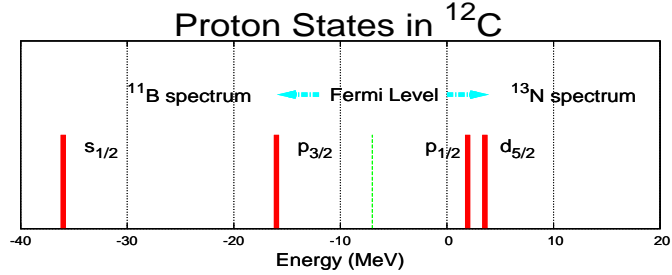}}
\caption{\label{F7} Single particle energies of  $^{12}$C.
The particle states (above the Fermi level) $1p_{1/2}$, and $1d_{5/2}$
 are the lowest states in $^{13}$N, while the hole states  (below the Fermi level)
  $1s_{1/2}$, and $1p_{3/2}$ are the lowest levels (in inverted order) in $^{11}$B.}
\end{figure}

\begin{figure}[th]
{\includegraphics[width=10.cm,height=6.cm]{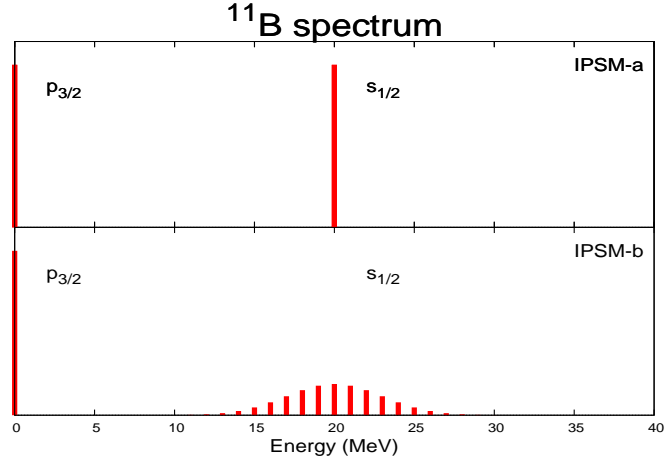}}
\caption{\label{F8} Schematical representation of
 the  $1s_{1/2}$, and $1p_{3/2}$  levels  in $^{11}$B within the models
 IPSM-a (upper panel), and IPSM-b (lower panel).  Due
to the coupling between hole and particle states, the deeply bound hole states acquire a width
and is treated as continuum state in the  IPSM-b model.}
\end{figure}

Before proceeding a few words should be said on the s.p.e.
of the emitted nucleon $N=n,p$  in the two-nucleon decay $\Lambda n\go nN$, which in
the first section
has been denoted by $\varepsilon_N^\downarrow$.  Let us consider the decay
$^{12}_\Lambda $C$\go^{10}$B$+\Lambda+p$, as one example.
The s.p.e. of  $^{12}$C  are displayed in Fig. \ref{F7}.
In the pure IPSM the particle states (above the Fermi level) $1p_{1/2}$, and $1d_{5/2}$
 are the lowest states in $^{13}$N, while the hole states  (below the Fermi level)
  $1s_{1/2}$, and $1p_{3/2}$ are the lowest levels (in inverted order) in $^{11}$B, as illustrated
in the upper panel in Fig. \ref{F8}. The $1s_{1/2}$ orbital is separated  from the
$1p_{3/2}$ state by approximately $23$ MeV, which
is enough to break the $10$ particle system of $^{10}$B, where the energy of the last
excited state amounts to $\sim 16.5$ MeV.
In fact, a single-particle state $\ket{j_N}$ that is  deeply bound in the
hypernucleus,  after the NMWD can become a highly excited hole-state
$\ket{j^{-1}_N}$ in the continuum of the residual nucleus.
There  it suddenly mixes up with more complicated configurations
(2h1p, 3h2p, \dots excitations, collective states, \etc.)
spreading its strength in a relatively wide energy interval~\cite{Ma85}, as schematically
represented in the lower panel in in Fig. \ref{F8}.
\footnote{One should keep in mind that the mean life a $\Lambda$ hyperon is
$\tau_\Lambda= 2.63 \times 10^{-10}$ s, while the strong interaction times are
of the order of $10^{-21}$ s.}
Although  the detailed structure and fragmentation
of hole states are still not well known, the
exclusive knockout reactions provide a wealth of
information on the structure of single-nucleon states
of nuclei. Excitation energies  and widths of
proton-hole states were systematically measured with
quasifree $(p, 2p)$ and $(e, e' p)$ reactions, which
revealed the existence of inner orbital shells in
nuclei~\cite{Ja73,Fr84,Be85,Le94,Ya96,Ya01,Yo03,Ya04,Ko06}.

Therefore, the following two approaches  for the final states $\ket{\alpha_NJ_F}$
will be examined within the IPSM:

\subsection{IPSM-a}

Here, we completely ignore the residual interaction and, consequently, the only
states $\ket{\alpha_NJ_F}$ giving a nonzero result in Eq. \rf{4.3} and therefore
contributing to Eq. \rf{3.8} are those obtained by the weak coupling,  and
properly antisymmetrizing,  of the one  hole (1h) states $\ket{j^{-1}_N}$
to the core ground-state $\ket{J_C}$.  Then,
\begin{equation}
\ket{\alpha_N J_F} \mapsto \ket{j_NJ_F} \equiv \ket{(J_C,j^{-1}_N)J_F}\,,
\quad \mathrm{and} \quad
E_{\alpha_N J_F} \mapsto E_{j_N} \equiv  E_C - \epsilon_{j_N} \,,
\label{4.3}\end{equation}
where $\epsilon_{j_N}$ is the single-particle energy of state $j_N$, and
 the liberated energy in Eq. \rf{4.1} becomes
\be \Delta_{\alpha_N J_F} \mapsto \Delta_{j_N} = \Delta +
\epsilon_{j_\Lambda} + \epsilon_{j_N}. \label{4.4}\ee

As an illustration, in the case of $^{28}_\Lambda$Si the model space contains
four single-particle states, both for protons and for neutrons ($\rn_p=\rn_n=4$), namely,
$1s_{1/2}$, $1p_{3/2}$, $1p_{1/2}$ and $1d_{5/2}$.
Thus,  if the core state is $\ket{J_C}=\ket{1d_{5/2}n^{-1}}$,
the final states $\ket{(J_C,j^{-1}_N)J_F}$ are constructed by creating  two holes in the
$^{28}$Si nucleus, and read:
\br
\begin{array}{ll}
\underline{_\Lambda^{28}{\rm Si}\rightarrow nn+{^{26}{\rm Si}}}&
\underline{_\Lambda^{28}{\rm Si}\rightarrow np+ {^{26}{\rm Al}}}\\\\
\ket{(1d_{5/2}n^{-1})^2;0,2,4} &
\ket{(1d_{5/2}n^{-1}1d_{5/2}p^{-1});0,1,2,3,4,5}\\
\ket{1d_{5/2}n^{-1}1s_{1/2}n^{-1};2,3} &
\ket{1d_{5/2}n^{-1}1s_{1/2}p^{-1};2,3}\\
\ket{1d_{5/2}n^{-1}1p_{1/2}n^{-1};2,3} &
\ket{1d_{5/2}n^{-1}1p_{1/2}p^{-1};2,3}\\
\ket{1d_{5/2}n^{-1}1p_{3/2}n^{-1};1,2,3,4} &
\ket{1d_{5/2}n^{-1}1p_{3/2}p^{-1};1,2,3,4}.
\end{array}
\label{4.5}\er

After making  the substitution \rf{4.4} in  Eqs. \rf{3.1}-\rf{3.28}
  one  can perform the summation on ${\alpha_N J_F}$
for each single-particle state $j_N$, as done
in~\cite[Eqs. (11), (12), (13)]{Kr03}, and do
\br
\mathcal{F}_{\alpha_NJ_F}(pP) \mapsto\mathcal{F}_{j_N}(pP)&=&\sum_{ J=|j_N-1/2|}
^{J=j_N+1/2}F^j_{NJ}\sum_{SlL\lambda T}|\M(plPL\lambda SJ\T;{j_\Lambda j_N J\tLN})|^2,
\nn\\
\label{4.6}\er
with the  $F^j_{NJ}$
are defined as
\br
F^j_{NJ}&=&\hat{J_I}^{-2}\sum_{J_F} |\Bra{J_I}\left( a_{j_N}^\dag
a_{j_\Lambda }^\dag\right)_{J}\Ket{J_F}|^2
\nn\\
&=&\hat{J}^{2}
\sum_{J_F}\sixj{J_C}{J_I}{j_\Lambda}{J}{j_N}{J_F}^2  |\Bra{J_C}a_{j_N}^\dag\Ket{J_F}|^2.
 \label{4.7}
\er
The general formula \rf{3.6}, \rf{3.11}, \rf{3.15}, and \rf{3.23}, read now
\br
\Gamma_{N}&=&
\frac{16M^3}{\pi}\left(\frac{A-2}{A}\right)^{3/2}\int_{0}^{\Delta} d\e_P
\sum_{j_N}
\sqrt{\e_P(\Delta_{j_N}-\e_P)}\mathcal{F}_{j_N}(pP),
\label{4.8}\er
\begin{equation}
\Gamma_N =
\frac{4M^3}{\pi}\sqrt{A(A-2)^3}\int_0^\Delta  dE_{nN}
\sum_{j_N}
\sqrt{(\Delta_{j_N}-E_{nN})(E_{nN}-\Delta_{j_N}')}
\mathcal{F}_{j_N}(pP),
\label{4.9}\ee
  \br
\Gamma_{N}&=&
\frac{2M}{\pi}\sqrt{\frac{A-2}{A}}\int_{0}^{P_\Delta}dP
\sum_{j_N}
P^2\sqrt{P^2_{\Delta_{j_N}}-P^2}\mathcal{F}_{j_N}(pP),
\label{4.10}\er
and
\br
\Gamma_N&=&(A-2)\frac{8M^3}{\pi}\sum_{j_N}  \int _{-1}^{+1}d \cos\theta_{nN}
\int_0^{E_{j_N}^{max}} dE_{N}
\sqrt{\frac{E_{N}}{E'}}E_n\mathcal{F}_{j_N}(pP).
 \label{4.11}\er
The meaning of all other quantities is self-evident from the initial expresions.

\subsection{IPSM-b}

Formally, one starts from the unperturbed basis $\ket{i_N J_F}_0$ with
$i_N=1,2,\dots \rn_N,\rn_N+1,\dots $, where for $i_N\le \rn_N$ we have the same
simple doorway states $\ket{j_NJ_F}$ in Eq. \rf{4.4}
(listed in Eq. \rf{4.5} for $^{28}_\Lambda$Si), while for $i_N \ge \rn_N+1$ we
have more complicated bound configurations
(such as $3h1p$, $4h2p$, \dots in the case of $^{28}_\Lambda$Si) as
well as those including unbound single-particle states in the continuum.
As in Ref.~\cite{Ma85}, the  perturbed  eigenkets  $\ket{\alpha_N J_F}$ and
eigenvalues $E_{\alpha_N J_F}$ are obtained by diagonalizing the matrix
$_0\bra{i_N J_F}H\ket{i'_N J_F}_0 $ of the exact Hamiltonian $H$:
\begin{equation}
\bra{\alpha_N J_F}H\ket{\alpha'_N J_F } =
E_{\alpha_N J_F} \, \delta_{\alpha_N \alpha'_N}
\label{4.12}
\end{equation}
with
\begin{eqnarray}
\ket{\alpha_N J_F} &=& \sum_{i_N=1}^\infty C_{i_N}^{\alpha_NJ_F} \ket{i_N J_F}_0
\nonumber \\
&=& \sum_{j_N} C_{j_N}^{\alpha_NJ_F} \ket{j_NJ_F}
+ \sum_{i_N=\rn_N+1}^\infty C_{i_N}^{\alpha_NJ_F} \ket{i_N J_F}_0 \,.
\label{4.13}
\end{eqnarray}
It is easy to see that only the ket $\ket{j_NJ_F}$ in the expansion \rf{4.13}
will contribute to the matrix element
$\bra{J_C}|a^\dagger_{j_N}|\ket{\alpha_N J_F}$ in Eq. \rf{3.2}.
Therefore, the  Eq. \rf{3.6} takes the form
\br
\Gamma_{N}&=&
\frac{16M^3}{\pi}\left(\frac{A-2}{A}\right)^{3/2}\int_{0}^{\Delta} d\e_P
\sum_{j_N\alpha_NJ_F} |C_{j_N}^{\alpha_NJ_F}|^2
\sqrt{\e_P(\Delta_{\alpha_NJ_F}-\e_P)}\mathcal{F}_{j_NJ_F}(pP),
\nn\\
\label{4.14}\er
where
\begin{eqnarray}
\mathcal{F}_{j_NJ_F}(pP)&=& \hat{J}_I^{-2} \sum_{lL\lambda SJT}
\left| \mathcal{M}(pPlL\lambda SJT;j_N)
\bra{J_I}|\left(a^\dagger_{j_N} a^\dagger_{j_\Lambda}\right)_J|\ket{j_NJ_F}
\right|^2 \,.
\label{4.15}
\end{eqnarray}

To evaluate the amplitudes $C_{j_N}^{\alpha_NJ_F}$  one would have to choose
the appropriate Hamiltonian $H$ and the unperturbed basis $\ket{i_N J_F}_0$,
and solve the eigenvalue problem \rf{4.12}.
We will not do this here. Instead, we  make a phenomenological
estimate~\cite{Ba08}. First, because of the high density of states, we will convert the
discrete energies $\Delta_{\alpha_NJ_F}$ into the  continuous variable
$\ve$, and the discrete  sum on ${\alpha_N}$ into an integral on $\ve$, \ie
\be
\Delta_{\alpha_NJ_F} \to \varepsilon \,, \quad
\sum_{\alpha_NJ_F} |C_{j_N}^{\alpha_NJ_F}|^2  \to
\sum_{J_F} \int_{-\infty}^{\infty} |C_{j_NJ_F}(\varepsilon)|^2
\rho_{J_F}(\varepsilon) d\varepsilon  \,,
\label{4.16}\ee
where $\rho_{J_F}(\varepsilon)$ is the density of perturbed states with
angular momentum $J_F$. In this way the Eq. \rf{4.14} becomes
\br
\Gamma_{N}&=&
\frac{16M^3}{\pi}\left(\frac{A-2}{A}\right)^{3/2}\int_{0}^{\Delta}d\e_P
\sum_{j_NJ_F} \int_{-\infty}^{\infty}d\varepsilon\rP_{j_NJ_F}(\varepsilon)
\e_P \sqrt{\e_P(\varepsilon-\e_P)}\mathcal{F}_{j_NJ_F}(pP),
\nn\\
\label{4.17}\er
where
\begin{equation}
\rP_{j_NJ_F}(\varepsilon) = |C_{j_NJ_F}(\varepsilon)|^2 \rho_{J_F}(\varepsilon)
\label{4.18}
\end{equation}
is called the \emph{strength function} \cite{Ma85,Fraz96,Sarg00} and represents
the probability of finding the configuration
$\ket{j_NJ_F} \equiv \ket{(J_C,j_N^{-1})J_F}$ per unit energy interval.
Moreover, the Eq. \rf{3.9} is substituted by
 \be
P=2\sqrt{\frac{A-2}{A}M\e_P},\hspace{0.5cm}p=\sqrt{M (\varepsilon-\e_P)},
 \label{4.19}\ee
and the condition $\varepsilon\ge\e_P$
has to be fulfilled throughout the $\varepsilon$ integration. It is convenient
to introduce the averaged strength function
\begin{equation}
\rP_{j_N}(\varepsilon) = \frac{1}{\dim(j_NJ_C)}
\sum_{J_F=|J_C-j_N|}^{J_C+j_N} \rP_{j_NJ_F}(\varepsilon) \,,
\label{4.20}
\end{equation}
where
\begin{equation}
\dim(j_NJ_C) = \left\{
\begin{array}{lcl}
2j_N+1&\mathrm{for}&j_N \le J_C \,,
\\
2J_C+1&\mathrm{for}&J_C < j_N \,.
\end{array}
\right.
\label{4.21}
\end{equation}
This allows to simplify Eq.~\rf{4.17} by making the approximation
$\rP_{j_NJ_F}(\varepsilon) \approx \rP_{j_N}(\varepsilon)$ to get
\br
\Gamma_{N}&=&
\frac{16M^3}{\pi}\left(\frac{A-2}{A}\right)^{3/2}\int_{0}^{\Delta}d\e_P
\sum_{j_N} \int_{-\infty}^{\infty}d\varepsilon\rP_{j_NJ_F}(\varepsilon)
\e_P \sqrt{\e_P(\varepsilon-\e_P)}\mathcal{F}_{j_N}(pP).
\nn\\
\label{4.22}\er

The IPSM-a results would be recovered if one made the further
approximation
\be
{\rm P}_{j_N}(\ve)=\delta(\ve-\Delta_{j_N}).
\label{4.23}\ee
Here, in IPSM-b, the $\delta$-functions \rf{4.23} will be used for the
strictly stationary states, while for the fragmented hole states we will use
Breit-Wigner distributions,
\be
{\rm P}_{j_N}(\ve)=\frac{2\gamma_{j_N}}{\pi}
\frac{1}{\gamma_{j_N}^2+4(\ve-\Delta_{j_N})^2},
\hspace{1cm}
\int^\infty_{-\infty}{\rm P}_{j_N}(\ve)d\ve=1,
\label{4.24}\ee
where  $\gamma_{j_N}$ are the widths of the resonance centroids at energies
$\Delta_{j_N}$ (see~\cite[Eq.(2.11.22)]{Ma85}).
One proceeds similarly with the Eqs. \rf{3.11}, \rf{3.15}, and \rf{3.23}.
 It turns out that the expressions within the IPSM-b  can be obtained from
those of IPSM-a through the replacements:
\be
\Delta_{j_N} \mapsto \ve,\hspace{.5cm}\mbox{and}\hspace{.5cm}
\sum_{j_N} \cdots \mapsto
\sum_{j_N} \int_{-\infty}^{+\infty} d\ve {\rm P}_{j_N}(\ve) \cdots .
\label{4.25}\ee

To evaluate the transition rates $\Gamma_N$  we need to
know the spectroscopic factors $F^j_{NJ}$ given by \rf{4.7}, which
 depend on the angular momenta $J_C$ and $J_I$,  experimental values of which are given
 in Table \ref{T1}.
It is also necessary to choose between the $jj$ and $LS$
couplings. As in the previous work~\cite{Kr03} (see Table I) we used here the $jj$-coupling,
which is extensively used in nuclear physics  in view of large spin-orbit
splitting. In fact,  the experimental $p_{1/2}-p_{3/2}$ energy difference
is $\approx 6$ MeV around $^{16}$O. The resulting spectroscopic factors are shown in
 Table \ref{T2}.

\begin{table}
\begin{tabular}{|c|ccccc ccccc ccc|}
\hline Nucleus &$^{3}_\Lambda$H &$^{4}_\Lambda$He &$^{4}_\Lambda$H
&$^{5}_\Lambda$He &$^{7}_\Lambda$Li &$^{9}_\Lambda$Be
&$^{11}_\Lambda$B &$^{12}_\Lambda$C &$^{13}_\Lambda$C
&$^{15}_\Lambda$N &$^{16}_\Lambda$O &$^{17}_\Lambda$O
&$^{28}_\Lambda$Si\\
\hline
  $J_C$  &$1$ &$1/2$ &$1/2$ &$0$&$1$  &$0$  &$3$ &$3/2$&$0$&$1$&$1/2$&$0$&$5/2$\\ \hline
$J_I$  &$1/2$ &$0$ &$1$ &$1/2$ &$1/2$ &$1/2$ &$5/2$
&$1$&$1/2$&$3/2$&$1$&$1/2$&$2$\\\hline
\end{tabular}
\caption{Experimental values of  core spin $J_C$, and  initial
spin $J_C$ for several $\Lambda$-hypernuclei.}
\label{T1}
\end{table}
\begin{table}
\begin{tabular}{|cc|ccccccccccccc|}
\hline
$j$&  $NJ$
&$^{3}_\Lambda$H&$^{4}_\Lambda$He&$^{4}_\Lambda$H &$^{5}_\Lambda$He
&$^{7}_\Lambda$Li&$^{9}_\Lambda$Be&$^{11}_\Lambda$B&$^{12}_\Lambda$C
&$^{13}_\Lambda$C&$^{15}_\Lambda$N&$^{16}_\Lambda$O&$^{17}_\Lambda$O&$^{28}_\Lambda$Si\\
\hline
  s$_{1/2}$& $n0$ &$3/2$ &$2$ &$1$ &$1$ &$1$&$1$ &$1$ &$1$&$1$&$1$&$1$&$1$&$1$\\
            &$n1$ &$1/2$ &$0$ &$3$&$3$ &$3$ &$3$ &$3$ &$3$&$3$&$3$&$3$&$3$&$3$\\ 
            &$p0$ &$3/2$ &$1$ &$2$ &$1$ &$1$ &$1$&$1$ &$1$&$1$  &$1$&$1$&$1$&$1$\\ 
            &$p1$ &$1/2$ &$3$ &$0$ &$3$ &$3$ &$3$&$3$ &$3$ &$3$ &$3$&$3$&$3$&$3$\\ 
\hline
$p_{3/2}$  &$n1$ &$-$ &$-$ &$-$   &$-$ &$5/2$&$3$&$13/2$&$7$ &$6$&$6$&$6$&$6$&$6$\\
           &$n2$&$-$&$-$&$-$&$-$&$3/2$&$5$&$11/2$&$5$&$10$&$10$&$10$&$10$&$10$\\ 
           &$p1$&$-$&$-$&$-$&$-$&$5/2$&$3$&$13/2$&$6$&$6$&$6$&$6$&$6$&$6$\\ 
           &$p2$&$-$&$-$&$-$&$-$&$3/2$&$5$&$11/2$&$10$&$10$&$10$&$10$&$10$&$10$\\
\hline
$p_{1/2}$  &$n0$ &$-$ &$-$ &$-$ &$-$&$-$ &$-$  &$-$ &$-$&$-$ &$0$&$0$&$1$&$1$\\
            &$n1$ &$-$&$-$&$-$&$-$&$-$ &$-$&$-$&$-$&$-$&$2$&$2$&$3$&$1$\\ 
            &$p0$ &$-$&$-$&$-$&$-$&$-$ &$-$&$-$&$-$&$-$&$0$&$1$&$1$&$1$\\ 
            &$p1$ &$-$&$-$&$-$&$-$&$-$&$-$&$-$&$-$ &$-$&$2$&$3$&$3$&$3$\\ 
\hline
$d_{5/2}$  &$n2$ &$-$&$-$&$-$ &$-$ &$-$ &$-$ &$-$  &$-$ &$-$&$-$ &$-$&$-$&$16$\\
           &$n3$&$-$&$-$&$-$&$-$&$-$&$-$&$-$&$-$ &$-$&$-$&$-$&$-$&$14$\\ 
           &$p2$&$-$&$-$&$-$&$-$&$-$&$-$&$-$&$-$&$-$&$-$ &$-$&$-$&$15$\\ 
           &$p3$&$-$&$-$&$-$&$-$&$-$&$-$&$-$&$-$&$-$&$-$&$-$ &$-$&$21$\\
\hline
\end{tabular}
\caption{Spectroscopic factors  $F^j_{NJ}$ multiplied by $2j+1$.}
\label{T2}
\end{table}

\section{NMWD Spectra}
The transition probability densities $S_{nN}(\e_{P})$, $S_{nN}(E_{nN})$
$S_{nN}(P)$, $S_{nN}(\cos\theta_{nN})$, and  $S_N(E_{N})$, can now be obtained
by performing derivatives on $\e_P$, $E_{nN}$, $P$, $E_{N}$, and $\cos\theta_{nN}$
in the appropriate equation for $\Gamma_N$, namely,
Eqs.~\rf{4.8},\rf{4.9},\rf{4.10}, \rf{4.11}, and \rf{4.11}, respectively.

\subsection {Effects of the deeply bound hole states}

In Ref. \cite{Ba08} we have studied the effects of the deeply bound hole states on the correlation spectra
$S_{nN}(E_{nN})$ of several hypernuclei.
The Fig.  \ref{F9}  shows the
normalized energy spectra  $ S_{np}(E)/\Gamma_p$ for $^{4}_\Lambda$He, $^{5}_\Lambda$He,
$^{12}_\Lambda$C, $^{16}_\Lambda$O, and $^{28}_\Lambda$Si
hypernuclei, evaluated within the full OMEP, that comprises the
($\pi,\eta,K,\rho,\omega,K^*$) mesons. Quite similar results are obtained for the nn pair, \ie for
$ S_{nn}(E)/\Gamma_n$.
The s.p.e.'s for the strictly stationary hole states have been taken from
Wapstra and Gove's compilation~\cite{Wa71}, and those of the quasi-stationary
ones have been estimated from the studies of the quasi-free scattering
processes $(p,2p)$ and  $(e,e'p)$
\cite{Ja73,Fr84,Be85,Le94,Ya96,Ya01,Yo03,Ya04,Ko06}.

\begin{figure}[h]
   \leavevmode
   \epsfxsize = 14cm
     \epsfysize = 9cm
\epsffile{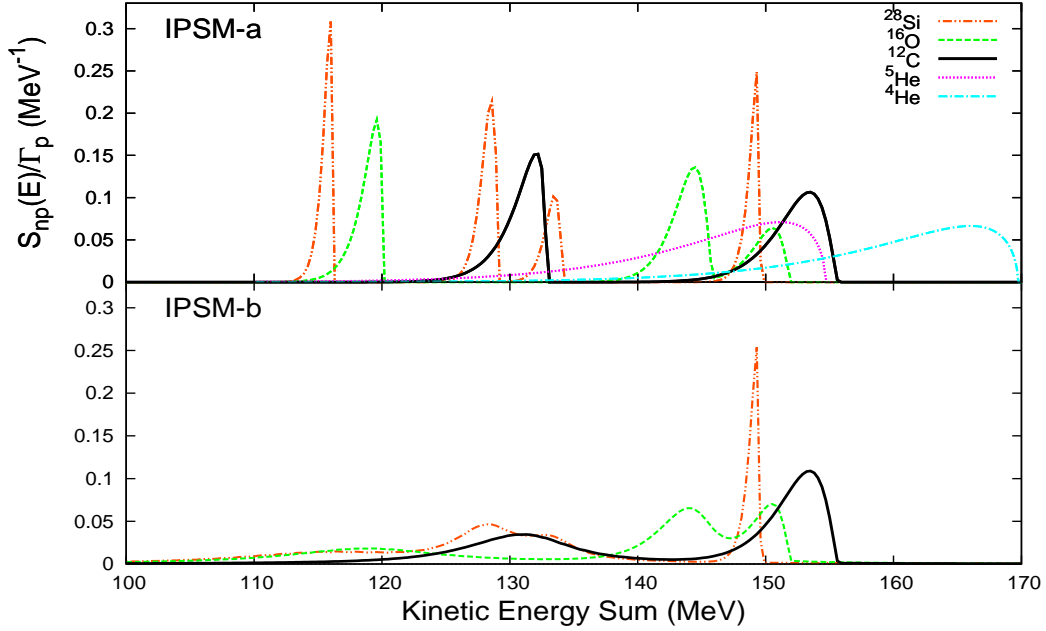}
\caption{\footnotesize
(Color online) Normalized energy spectra  $ S_{np}(E)/\Gamma_p$ for
$^{4}_\Lambda$He, $^{5}_\Lambda$He, $^{12}_\Lambda$C, $^{16}_\Lambda$O, and
$^{28}_\Lambda$Si hypernuclei for the full OMEP
obtained within the approaches IPSM-a (upper panel) and  IPSM-b (lower panel).
For the $s$-shell hypernuclei, only the IPSM-a approach has been used.}
\label{F9}\end{figure}

The two IPSM approaches exhibit some quite important differences:
\bit
 \item[a)] {\em IPSM-a}:  The spectra cover the
energy region $110$ MeV $<E< 170$ MeV and contain one or more
peaks, the number of which is equal to the number of shell-model orbitals
$ 1s_{1/2}, 1p_{3/2}, 1p_{1/2}, 1d_{5/2}, 2s_{1/2}, 1d_{3/2} \cdots$
that are either fully or partly occupied in $\ket{J_C}$.
Before including the recoil, all these peaks would be just spikes
at the liberated energies $\Delta_{j_N}$, as can be seen from \rf{3.4} setting
$E_r=0$. With the recoil effect, they behave as
\be
S_{nN}(E \cong\Delta_{j_N})\sim \sqrt{(\Delta_{j_N}-E)(E-\Delta'_{j_N})}
e^{-M(A-2)(\Delta_{j_N}-E)b^2},
\label{5.1}
\ee
and develop rather narrow widths $\sim [b^2M(A-2)]^{-1}$, where  $b$ is
the harmonic oscillator  size parameter, which  has been  taken from
Ref.~\cite{It02}.
These  widths go  from $\cong 3$ MeV for $^{28}_\Lambda$Si to $\cong 20$ MeV
for $^{4}_\Lambda$He, as indicated in the upper panels of the just mentioned
figures.
\item[b)]{\em IPSM-b}:
In the lower panels of the same figures are shown the results obtained
when the recoil is  convoluted with  the  Breit-Wigner distributions
\rf{4.24} for the strength functions of the fragmented deep hole states.
The widths $\gamma_{j_N}$ have been estimated from
Refs.~\cite{Ja73,Fr84,Be85,Ma85, Le94,Ya96,Ya01,Yo03,Ya04,Ko06},
and in particular from~\cite[Fig.11]{Ja73} and~\cite[Table 1]{Ya96},
with  results:
$\gamma_{{1s_{1/2}}}=9$ MeV in $^{12}_\Lambda$C,
$\gamma_{{1s_{1/2}}}=14$ MeV
and $\gamma_{{1p_{3/2}}}=3$ MeV in $^{16}_\Lambda$O,
\footnote{The  ${3/2}_1^-$ peak is at  $6.32$ MeV, but  small amounts of
the $p_{3/2}$ strength are also fragmented to the states of
$9.93$ MeV and $10.7$ MeV~\cite{Ko06}.} and
$\gamma_{{1s_{1/2}}}=16$ MeV
and $\gamma_{{1p_{3/2}}}=\gamma_{{1p_{1/2}}}=5$ MeV in $^{28}_\Lambda$Si,
both for protons and neutrons.
One sees that, except for the ground states, the narrow peaks engendered by the
recoil effect become now pretty  wide  bumps.
\eit

We feel that the above  rather rudimentary  parameterization  could be
realistic enough for a qualitative discussion of the kinetic energy sum spectra.
A more accurate  model should be probably  necessary for  a full quantitative
study and comparison with data.

\subsection{Interpretation of   BNL experiment
E788 on  $^4_\Lambda$He}

Particularly interesting is the Brookhaven National Laboratory experiment
E788 on $^{4}_\Lambda$He, performed by Parker \etal~\cite{Par07},
which highlighted that the effects of the Final State Interactions (FSI) on
the one-nucleon induced decay, as well as the contributions of the two-nucleon
induced decays, $\Lambda NN \go nNN$,  could be very small in this case,
if any.

Therefore one might hope that the IPSM could be an adequate framework to account
for the NMWD spectra of this hypernucleus. This has been done in Ref.~\cite{Bau09}
by employing the  $\pi+ K$ exchange potential,
 with soft cutoffs ($\Lambda_\pi= 0.7$ GeV and  $\Lambda_K=0.9$ GeV), which
 is capable of accounting for the experimental values
related to $\Gamma_p$ and $\Gamma_n$
in all three $^4_\Lambda$H, $^4_\Lambda$He, and $^5_\Lambda$He hypernuclei~\cite{Bau09}.
It is labelled as SPKE model and  is not very different from the PKE model used by
Sasaki \etal~\cite{Sa02}.

The transition probability densities $S_{N}(E)$, $S_{nN}(E)$, and
$S_{nN}(\cos\theta)$
contain the same dynamics, but involve different
phase-space kinematics for each case.
In particular, the proton spectrum
$S_{p}(E)$ is related with the expected number
of protons  $d{\rm N}_p(E)$   detected within the energy
interval $dE$ through the relation
\be
\frac{d{\rm N}_p(E)}{dE}=C_p(E)S_p(E),
\label{5.2}\ee
 where
$C_p(E)$ depends on the proton experimental environment and
includes all quantities and effects not considered in $S_p(E)$,
such as  the  number of produced hypernuclei,
the detection efficiency and acceptance, \etc.
In experiment E788, after correction for
acceptance, the remaining $C_p(E)$ factor is approximately energy-independent
in the region beyond the detection threshold, $E^0_p$~\cite{Par08}.
In what follows, we will always compare our predictions
with the experimental spectra that have been corrected for
acceptance and take into account the detection threshold.
Thus we can write, for the expected number of detected protons above
this threshold,
  \be
\bar{{\rm N}}_{p}=\int_{E_p^0}^{E_p^{max}}
\frac{d{\rm N}_{p}(E)}{dE}dE=
\bar{C}_{p}\int_{E_p^0}^{E_p^{max}} S_{p}(E)dE=
\bar{C}_{p}\bar{\Gamma}_p.
\label{5.3}\ee
\begin{figure}
\includegraphics[width=0.8\linewidth]{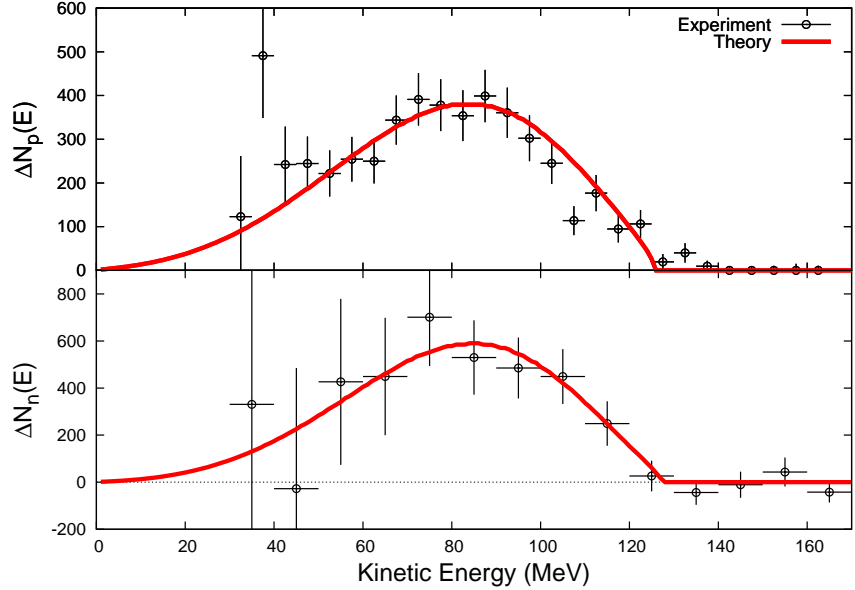}
\caption{\label{F10} Comparison between the experimental and theoretical
kinetic energy spectra for protons (upper panel) and neutrons (lower panel).
The data are acceptance corrected~\cite{Par08},
and the calculated results are obtained from Eqs. \rf{5.6} and \rf{5.7}.}
\end{figure}

 This allows us to rewrite \rf{5.2} in the form%
\footnote{A similar expression is valid for  the $\beta$-decay
 strength function (see, for instance, \cite[Eq. (5)]{Wi06}).}
 \be
\frac{d{\rm N}_p(E)}{dE}={\bar{{\rm N}}_p}\frac{S_p(E)}{\bar{\Gamma}_p}
\qquad (E > E_p^0).
\label{5.4}\ee
The  spectrum  $S_p(E)$ is normalized to the experimental one
by replacing $\bar{{\rm N}}_p$ in \rf{5.4} with the
acceptance-corrected number of actually observed protons,
\be
\bar{{\rm N}}_p^{exp}=\sum_{i=1}^{m} \Delta {\rm N}_p^{exp}(E_i),
\label{5.5}\ee
where $\Delta {\rm N}_p^{exp}(E_i)$ is the acceptance-corrected number of
protons  measured at energy $ E_i$ within a fixed energy bin $\Delta E_p$,
and $m$ is the number of bins beyond the detection threshold.
Thus, the quantity that we have to confront with data is
%
 \be
\Delta {\rm N}_p(E)=
{\bar{\rm N}}_p^{exp}\Delta E_p\frac{S_p(E)}{\bar{\Gamma}_p},
 \label{5.6}\ee
where the barred  symbols ($\bar{\rm N}_p^{exp}=4546$, and
$\bar{\Gamma}_p=  0.168$) indicate that the proton threshold  $E^0_p= 40$  MeV
\cite{Par08} has been considered in the numerical evaluation of the
corresponding quantities.
In contrast to  $\Delta {\rm N}_p^{exp}(E_i)$, $\Delta {\rm N}_p(E)$
is a continuous function of $E$.

\begin{figure}
\includegraphics[width=.8\linewidth]{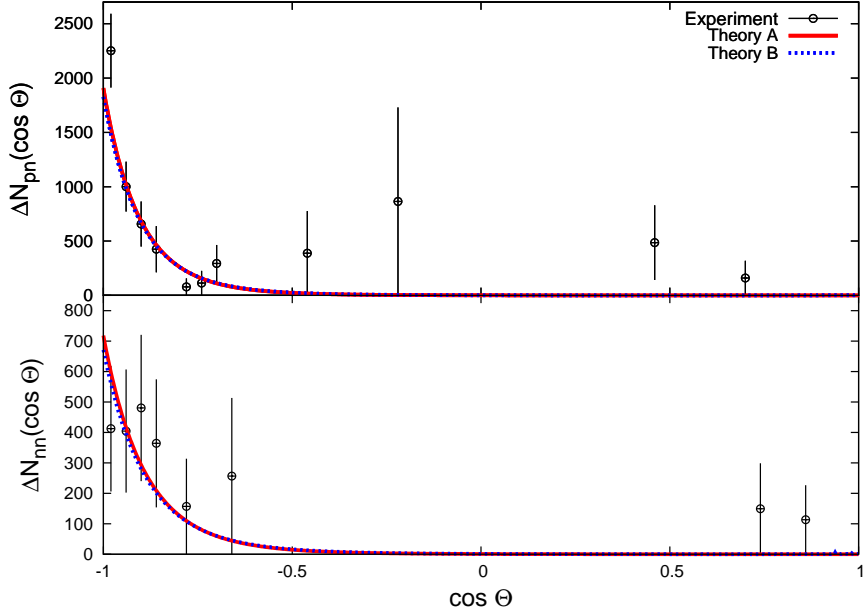}
\caption{\label{F11} Comparison between experimental opening angle
correlations for proton-neutron (upper panel) and neutron-neutron (lower panel)
pairs. The data $\widetilde{\Delta {\rm N}}_{nN}^{exp}(\cos\theta_i)$ are
acceptance corrected and do not  contain
events with  $E_N<30$ MeV~\cite{Par08}.
The theoretical results are obtained from  Eq. \rf{5.9},
with $\widehat{{\rm N}}_{nN}^{exp}$
only containing  events with  $\cos\theta_{nN}<-0.5$.
Two cases are presented: 1) Theory A, where
both the angular and the single kinetic energy cuts are taken into account,
and 2) Theory B,
where the cuts are not considered in the calculations.}
\end{figure}

As the one-proton (one-neutron) induced decay prompts the emission of an $np$
($nn$) pair, one has in the same way for the one-neutron spectrum
\be
\Delta {\rm N}_n(E)= {\bar{\rm N}}_n^{exp}\Delta E_n
\frac{S_p(E)+2S_n(E)}{\bar{\Gamma}_p+2\bar{\Gamma}_n}.
 \label{5.7}\ee
Here, ${\bar{\rm N}}_n^{exp}=3565$, and  $\bar{\Gamma}_p+2\bar{\Gamma}_n=0.198$
have been  evaluated with a neutron threshold of $ 30$  MeV \cite{Par08}.
In Fig. \ref{F10}, our results are compared  with the  measurements of
Parker \etal~\cite{Par07}.
A similar, but somewhat different, procedure is followed for the coincidence
spectra. The main difference arises from the fact that
the angular-correlation spectra, $\Delta {\rm N}_{nN}^{exp}(\cos\theta_i)$,
as well as the  kinetic energy sum data, $\Delta {\rm N}_{nN}^{exp}(E_i)$,
besides being  acceptance-corrected, were measured  with  detection thresholds
of $30$ MeV for both neutrons and protons.
More, in the selection of the kinetic energy sum data it
was also applied an angular cut of
$\cos\theta_{nN}<-0.5$.
In order to make the presentation simple,
the observables that comprise only the energy cuts, and those that
include both the energy and the angular cuts,
will be indicated by putting, respectively,  a tilde and a hat over the
corresponding symbols.

Thus, the number of $nN$ pairs measured in coincidence can be expressed as
\be
\widehat{{\rm N}}_{nN}^{exp}
=\sum_{i=1}^{k} \widetilde{\Delta {\rm N}}_{nN}^{exp}(\cos\theta_i)=
\sum_{i=1}^{l} {\widehat{\Delta{\rm N}}}_{nN}^{exp}(E_i),
\label{5.8}\ee
where  the angular bins with $\cos\theta_i>-0.5$ are excluded from the first
summation.
The $\widetilde{\Delta {\rm N}}_{nN}^{exp}(\cos\theta_i)$ and
${\widehat{\Delta{\rm N}}}_{nN}^{exp}(E_i)$ data
should be compared, respectively, with
\be
\widetilde{\Delta {\rm N}}_{nN}(\cos\theta)=
\widehat{{\rm N}}_{nN}^{exp}\Delta\cos\theta_{nN}
\frac{\widetilde{S}_{nN}(\cos\theta)}{\widehat{{\Gamma}}_N},
\label{5.9}\ee
and
\be
\widehat{\Delta {\rm N}}_{nN}(E)=\widehat{{\rm N}}_{nN}^{exp}\Delta E_{nN} \frac{\widehat{S}_{nN}(E)}{\widehat{{\Gamma}}_N}.
\label{5.10}\ee
Here, from Ref.~\cite{Par08}  $\widehat{{\rm N}}_{np}^{exp}=4821$,
$\widehat{{\rm N}}_{nn}^{exp}=2075$,
$\Delta\cos\theta_{nN}=0.04$ and $\Delta E_{nN}=10$ MeV,
while
$\widehat{\Gamma}_p=0.1709$ and
$\widehat{\Gamma}_n=0.0113$.
These  results (Theory A) are compared with the E788 data in
Figs. \ref{F11} and \ref{F12}.
For completeness, in the same figures are also shown the results for
$\widetilde{S}_{nN}(\cos\theta)\go{S}_{nN}(\cos\theta)$,
$\widehat{S}_{nN}(E)\go {S}_{nN}(E)$ and
$\widehat{\Gamma}_N\go {\Gamma}_N$,
\ie when no energy and angular
cuts are considered in the theoretical evaluation, and ${\Gamma}_p=0.1793$ and
${\Gamma}_n=0.0122$  (Theory B).

\begin{figure}[h]
\includegraphics[width=0.8\linewidth]{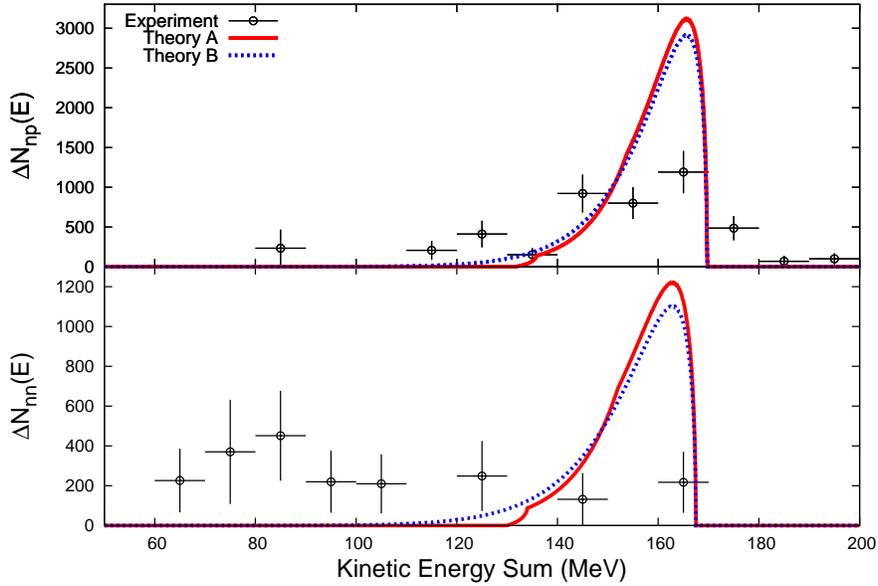}
\caption{\label{F12} Comparison between experimental kinetic energy sum
spectra for proton-neutron (upper panel) and neutron-neutron (lower panel)
pairs.
The data $\widehat{\Delta {\rm N}}_{nN}^{exp}(E_i)$ are acceptance corrected
and only contain events with  $E_N>30$ MeV and
$\cos\theta_{nN}<-0.5$~\cite{Par08}.
The theoretical results are obtained from  Eq. \rf{5.10},  and
two cases are shown: 1) Theory A, where
both cuts are taken into account, and 2) Theory B,
where the cuts are not considered in the calculations.}
\end{figure}

We conclude that the overall agreement between the  measurements of
Parker \etal~\cite{Par07} and the present calculations is quite satisfactory,
although we are not considering contributions coming from the two-body
induced decay, $\Lambda NN\go nNN $,  nor from the rescattering of the nucleons
produced in the one-body induced decay, $\Lambda N\go nN$.
However, before ending the discussion we would like to point out that:
\bnu
\item As expected,  the theoretical spectrum $\Delta {\rm N}_{p}(E)$,
shown in the upper panel of  Fig. \ref{F10}, is
peaked around $85$ MeV, corresponding to the half of the $Q$-value
$\Delta_p=170$ MeV.
Yet, as the  single kinetic energy reaches rather abruptly its maximum value
$E^{max}_p=127$ MeV  (see Eq. \rf{3.28}), the proton spectrum shape is  not
exactly  that of a symmetric inverted bell.
Something quite analogous  happens in the case of neutrons, as can be seen in
the lower panel of  Fig. \ref{F10}.
The experimental data seem to behave in the same way.
To some extent, this   behavior of $\Delta {\rm N}_{p}(E)$ and
$\Delta {\rm N}_{n}(E)$  is akin to the   behavior  of the
$\Delta {\rm N}_{nN}(E)$, which  suddenly  collapse  at the Q-values.

\item There are no data at really low energies for the
proton case which would allow to exclude the FSI effects for sure, and
the neutron data for low energies are afflicted by large error
bars. However, there is no need to invoke these effects, nor those of
two-nucleon induced NMWD, to explain the data, as occurs in the proton
spectrum of $^{5}_\Lambda$He~\cite{Ag08}. This hints at a new puzzle in
the NMWD, but it is difficult to discern whether it is of experimental or
theoretical nature.

 \item The  calculated spectra  $\widetilde{\Delta {\rm N}}_{np}(\cos\theta)$
shown in the upper panel of Fig. \ref{F11},
are  strongly peaked near $\theta=180^o$,
which agrees with data fairly well. However, while it is found experimentally
that $28 \%$ of events occur  at  opening angles less than $120^o$,
theoretically we get that only $\lsim 2 \%$ of events appear in this angular
region. We find no explanation for this discrepancy.
Nevertheless, the  fact that not all events are concentrated at $\theta=180^o$,
is not necessarily indicative of the  contributions coming from the  FSI or the
$\Lambda NN\go nNN $ decay, as suggested  in Ref. \cite{Par07}.

\item
The calculated angular correlation
$\widetilde{\Delta {\rm N}}_{nn}(\cos\theta)$,  shown in
the lower panel of Fig. \ref{F11}, is quite similar to that of the
$pn$ pair; that is, its   back-to-back peak is very pronounced.
This  behavior  is not exhibited by the experimental distribution.
In addition, while  $11 \%$ of events are found experimentally
for $\cos\theta \ge -0.5$, in the calculation only $\lsim 3 \%$ of them appear
at these angles.
We feel however that, because of the poor statistics and large experimental
errors, one should not attribute major importance to such  disagreements.

\item
Both calculated kinetic energy sum distributions
$\widehat{\Delta {\rm N}}_{nN}(E)$,  shown in
Fig. \ref{F12}, present a bump at $\approx 160$ MeV, with a
width of $\approx 30$ MeV, which for protons agrees fairly well with
the experiment.
We would like to stress once more that the
spreading in strength here is totally normal even for a purely
one-nucleon induced decay.
The kink at  $\approx 130$ MeV within the Theory A comes from the angular cut,
and from this one can realize  that the ${nN}$ kinetic energy sum spectra
below this energy are correlated with the angular coincidence spectra
$\widetilde{\Delta {\rm N}}_{nN}(\cos\theta<-0.5)$.
The bump  observed  in the experimental $\widehat{\Delta {\rm N}}_{nn}(E)$
spectrum at $\approx 90$ MeV is not reproduced by the theory,
which may be indicative of
$nn$ coincidences originated from sources other than  $\Lambda n $ decays,
as already suggested in Ref.~\cite{Pa07}.
Another source for the difference between our model
calculation and the data may be traced to $np$ and $nn$ final state
interactions. Whereas in the former the intensity of this interaction
is reduced owing to the Coulomb repulsion felt by the proton, in
the latter the two neutrons may interact strongly and thus shift the
peak to lower kinetic energy sum.
 \enu

In summary,  to comprehend the recent   measurements in $^4_\Lambda$He,
we have outlined  for the one-nucleon induced NMWD spectra
a simple theoretical framework based on  the IPSM.
Once normalized to the transition rate,  all the spectra are tailored
basically by the kinematics of
the corresponding phase space, depending very
weakly on the dynamics
governing the $\Lambda N \to nN$ transition proper.
As a matter of fact, although not shown here, the normalized spectra calculated
with the full PSVE model are, for all practical purposes, identical
to those using the SPKE model, which we have amply discussed.
In spite of the simplicity of the approach,
a good agreement with data is obtained.
This might indicate that, neither  the FSI, nor the two-nucleon induced
decay processes play a significant role
in the $s$-shell, at least not for $^4_\Lambda$He.

\section{Outlook}

Before being detected the newborn nucleons in a NMWD  suffer  final state interactions
(FSI) with the nuclear environment, and consequently their two-nucleon $\Gamma_N$,
and  three-nucleon $\Gamma_2$ decay rates
are not observable from a quantum-mechanical point of view,  as recently pointed out
by Bauer and Garbarino~\cite{Ba10}. These FSI give rise to  emission of new
 secondary nucleons, that are counted by the detection systems together with the primordial ones,
 without being possible to distinguish ones from the others. The IPSM developed so far is  a simple
  fully quantum-mechanical formalism
for the theoretical investigation of  decay rates $\Gamma_N$ and their spectra. It  don't
describe  neither the decay rate $\Gamma_2$ nor the FSI.
Therefore, it is not surprising that this model does not reproduce well
 the  FINUDA experiment for the
$^{12}_\Lambda $C \cite{Ag08}, as shown in Fig. \ref{F13}, although it reproduces
 well \cite{Bau09} the BNL experiment for $^4_\Lambda$He
\cite{Par07}.

\begin{figure}[h]
\includegraphics[width=0.45\linewidth]{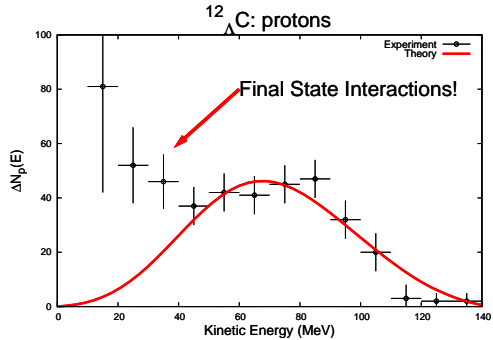}
\caption{\label{F13} Comparison between the experimental \cite{Ag08} and
theoretical~\cite{Ba10} kinetic energy spectra for protons from $^{12}_\Lambda$C decay. }
\end{figure}

At the time being we are studying  the
proton  kinetic energy spectra obtained in  the FINUDA  experiment  for
$^{5}_\Lambda$ He, $^{7}_\Lambda$Li, $^{9}_\Lambda$Be, $^{11}_\Lambda$B,
 $^{12}_\Lambda$C,  $^{15}_\Lambda$C, and $^{16}_\Lambda$O~\cite{Ag09,Ag10}.  We are comparing
 their results with the simple IPSM, with the purpose to quantify the contributions of the
 FSI and the three-nucleon emission. The same is being done for the
 recent KEK measurments on angular correlations and kinetic energy sum  of $np$ and $nn$
pairs~\cite{Ok04,Ka06,Ki06,Ki09}, as well as the c.m. momentum spectra
 in  $^{12}_\Lambda$C~\cite{Ki09}. Later on we will include the FSI,
a consistent treatment of which would require in general a genuine three-body approach
for the mutual interaction of the two emitted nucleons and the residual nucleus.
Presumably due to the enormous computational challenges, this has never been tackled in
the past.
 We are also planning to extend IPSM for the evaluation of
 the decay rate $\Gamma_2$ for the emission  of three primordial nucleons, which has been
 done so far only in the framework of the Fermi gas model~\cite{Al91,Ra94,Ba03,Bau109,Bau209}.

\begin{theacknowledgments}
 This work was partly supported by the Argentinian agency CONICET under
contract PIP 0377, and  by the Brazilian agencies FAPESP and CNPq.
We would like to thank Airton Deppman, Sergio Duarte, Eduardo Bauer, and
 Gianni Garbarino for very helpful discussions, and to Cl\'audio De Conti,
  Alejo C. Scarano, and
 Cecilia M. Krmpoti\'c for technical support.

\end{theacknowledgments}



\bibliographystyle{aipproc}   



\end{document}